\documentclass[runningheads]{cl2emult}

\usepackage{makeidx}  
\usepackage{graphicx} 
\usepackage{subeqnar} 
\usepackage{multicol} 
\usepackage{cropmark} 
\usepackage{eso}      
\makeindex            

\begin{document}
\title*{Data Mining in Astronomical Databases}
\toctitle{Data Mining in Astronomical Databases}
%
%
\titlerunning{Data Mining in Astronomical Databases}
%
\author{Kirk D. Borne\inst{1}}
\institute{Raytheon ITSS, NASA-GSFC, Code 631, Greenbelt, MD 20771}

\maketitle              

\begin{abstract}
A Virtual Observatory (VO) will
enable transparent and efficient access, search, retrieval, and
visualization of data across multiple data repositories, which are
generally heterogeneous and distributed.   Aspects of
data mining that apply to a variety of science user
scenarios with a VO are reviewed.  
\end{abstract}

\section{Science Requirements for Data Mining}
What is data mining and why is applicable to scientific research?  Data
mining is defined as {\it{an information extraction activity whose goal
is to discover hidden facts contained in databases}}.  Data mining has
taken the business community by storm and 
there is consequently now a vast array of resources and
research techniques available for exploitation by the scientific
communities.  It is useful therefore to examine a 
categorization of data mining thrusts and their sub-components, since
these are likewise applicable to the scientific exploration of large
astronomical databases.  Data mining is used to find patterns and
relationships in data by using sophisticated techniques to build models
-- abstract representations of reality. A good model is a useful guide
to understanding that reality and to making decisions.  There are two
main types of data mining models:  {\it{descriptive}} and
{\it{predictive}}.  {\it{Descriptive}} models describe patterns in
data and are generally used to create meaningful subgroups or
clusters.  {\it{Predictive}} models are used to forecast explicit
values, based upon patterns determined from known results.  

There is another differentiation of data mining into two categories
that we find particularly appropriate to knowledge discovery in
large astronomical databases:  {\it{event-based mining}} and
{\it{relationship-based mining}}.
At the risk of trivializing some fairly sophisticated techniques,
we classify event-based mining scenarios into four orthogonal categories:

\begin{itemize}

\item Known events / known algorithms -- use existing physical models
({\it{descriptive models}}) to locate known phenomena of interest
either spatially or temporally within a large database.

\item Known events / unknown algorithms -- use pattern recognition
and clustering properties of data
to discover new observational (in our case, astrophysical) relationships
among known phenomena.

\item Unknown events / known algorithms -- use expected
physical relationships ({\it{predictive models}}) among
observational parameters of astrophysical phenomena to predict
the presence of previously unseen events within a large complex database.

\item Unknown events / unknown algorithms -- use thresholds to identify
transient or otherwise unique (``one-of-a-kind'') events and therefore
to discover new phenomena.

\end{itemize}

\noindent
For relationship-based mining, we identify three classes
of association-driven scenarios that would find application in 
astronomical research:

\begin{itemize}

\item Spatial associations -- identify
events (astronomical objects) at the same location in the sky.

\item Temporal associations -- identify
events occurring during the same or related periods of time.

\item Coincidence associations -- use clustering
techniques to identify events that are co-located
within a multi-dimensional parameter space.

\end{itemize}

From this discussion, we thus derive a reduced set of 
science requirements for data mining that
correspond to the following exploratory approaches to
mining large databases\thinspace :
{\it{Object Cross-Identification}},
{\it{Object Cross-Correlation}},
{\it{Nearest-Neighbor Identification}}, and
{\it{Systematic Data Exploration}}.
(a)~``Object cross-identification'' refers to the classical problem of
connecting the source list in one catalog (or observation database) to
the source list in another, in order to derive new astrophysical
understanding of the cross-identified objects (e.g., gamma-ray burst
counterparts).  
(b)~``Object cross-correlation'' refers to the application
of ``what if'' scenarios to the full suite of parameters in a database
(e.g., identify distant galaxies as $U$-band dropouts in a color-color
scatter plot from the HDF survey).  
(c)``Nearest-neighbor identification'' refers to the general application
of clustering algorithms in multi-dimensional parameter space (e.g., 
finding the closest known population of young stars --
in the TW Hydrae association -- through their similar kinematics, 
X-ray emission, $H\alpha$, and Li abundance).
(d)~``Systematic data exploration'' refers to the application of the
broad range of event-based and relationship-based queries to a database
in the hope of making a serendipitous discovery of new objects or a 
new class of objects (e.g., finding new types of variable stars, such as
``bumpers'', in the MACHO database).

\section{User Scenarios for Mining Astronomical Databases}

It is well established that a significant fraction of all galaxies have
been involved in galaxy-galaxy interactions at
some time(s) in their past.   The rate of these interactions is not yet
well determined empirically.  In our first astronomical data mining
user scenario, we attempted an initial exploration of several on-line databases
in order to estimate the galaxy interaction rate.  We began by
exploring an on-line catalog of galaxies (available through NASA's ADC
= Astronomical Data Center):  the Updated Zwicky Catalog \cite{F1}.
This catalog identifies multiple-galaxy groupings, which
we used to reduce the full list of 19,000 galaxies to the set of 1800
multiples.  We then selected a very small sub-sample from this list to
conduct a proof-of-concept investigation.  We used existing catalog
visualization tools and archive linkage tools at the ADC to find all
possible NASA mission data and most of the all-sky survey data for these
selected objects \cite{K1}.  We then identified
characteristics in the optical images or in the IRAS fluxes or in the
X-ray emissions to verify that the associated multiple galaxy systems
are in fact (to high probability) bound groups (pairs, triples,
quartets, etc.).  The expectation that these small galaxy-galaxy
separations and other evidences for physical association do in fact
imply an on-going interaction was often confirmed through inspection of
the DSS (Digital Sky Survey) imagery, which showed signs of interaction
in many cases (e.g., distorted morphologies).  Thus, by applying
knowledge of astrophysical signatures of interactions, we were able to
explore multiple distributed databases in a coherent organized manner.  
We estimate
that the galaxy interaction rate in the local Universe is $\sim$8\%.

Among the exciting results of the COBE mission was the discovery 
of an extragalactic CIB (Cosmic Infrared Background; \cite{H2}).
There has been a storm of activity 
to identify the sources of the CIB and to understand 
what powers the strong IR emissions.  In our second data mining scenario,
we initiated a proof-of-concept search scenario for identifying
potential candidate contributors to the CIB.  Our approach is similar
to that of \cite{H1}, except that we are applying the
full power of on-line databases and linkages between these databases,
archives, and published literature.  Our search scenario involved
finding object cross-identifications among the IRAS Faint Source
Catalog and FIRST survey catalog, and then attempting to find those
commonly identified objects also within other databases, such as the
HST observation log.  In a very limited test sample of targets,
we did find one object in common among the HST-IRAS-FIRST databases: a
known hyperluminous infrared galaxy (HyLIRG) at
$z$=0.780 harboring an AGN, 
which was specifically imaged by HST because of its known
HyLIRG characteristics.  In this limited test scenario, we
did in fact find what we were searching for: a distant IR-luminous
galaxy that is a likely contributor to the CIB, or else a local analog
of the more distant objects that likely comprise the CIB.

Our two user scenarios (including Web screen shots) 
are presented at the following web site (under
the category ``How to use the ADC for scientific research projects'')~:~
{\tt{http://adc.gsfc.nasa.gov/adc/how\_to.html}} .

\clearpage
\addcontentsline{toc}{section}{Index}
\flushbottom
\printindex

\end{document}